\documentclass[conference]{IEEEtran}
\IEEEoverridecommandlockouts
\usepackage[hyphens]{url}
\usepackage{cite}
\usepackage{amsmath,amssymb,amsfonts}
\usepackage{algorithmic}
\usepackage{graphicx}
\usepackage{textcomp}
\usepackage{xcolor}
\usepackage{soul}
\usepackage[OT1]{fontenc}
\usepackage{paralist}
\usepackage{pdfcomment}     
\usepackage{pbalance}       
\usepackage{booktabs}

\hypersetup{
    colorlinks,
    linkcolor={red!50!black},
    citecolor={blue!50!black},
    urlcolor={blue!80!black}
}

\newcommand{\eg}{e.g.,\ }
\newcommand{\ie}{i.e.,\ }

\def\BibTeX{{\rm B\kern-.05em{\sc i\kern-.025em b}\kern-.08em
    T\kern-.1667em\lower.7ex\hbox{E}\kern-.125emX}}

\begin{document}

\bstctlcite{IEEEexample:BSTcontrol}

\title{Evaluating the Impact of Inter-cluster Communications in Edge Computing}

\author{\IEEEauthorblockN{Marc~Michalke,~Iulisloi~Zacarias,~Admela~Jukan}
\IEEEauthorblockA{\textit{Institute of Computer and Network Engineering,} \\
\textit{Technische Universit\"at Braunschweig}, Germany \\
\{m.michalke, i.zacarias, a.jukan\}@tu-bs.de}
\and
\IEEEauthorblockN{Kfir Toledo, Etai Lev-Ran}
\IEEEauthorblockA{\textit{IBM Research}, Israel \\
\{kfir.toledo@ibm.com, etail@il.ibm.com\}}


}


\maketitle


\begin{abstract}

Distributed applications based on micro-services in edge computing are becoming increasingly
popular due to the rapid evolution of mobile networks. While Kubernetes is the default framework
when it comes to orchestrating and managing micro-service-based applications in mobile networks,
the requirement to run applications between multiple sites at cloud and edge poses new
challenges. Since Kubernetes does not natively provide tools to abstract inter-cluster
communications at the application level, inter-cluster communication in edge computing is
becoming increasingly critical to the application performance. In this paper, we evaluate for
the first time the impact of inter-cluster communication on edge computing performance by using
three prominent, open source inter-cluster communication projects and tools, i.e., Submariner,
ClusterLink and Skupper. We develop a fully open-source testbed that integrates these tools in a
modular fashion, and experimentally benchmark sample applications, two of them centered around ML,
on their performance running in a multi-cluster edge computing system under varying networking
conditions. We experimentally analyze two classes of envisioned mobile applications, i.e., a)
industrial automation, b) vehicle decision drive assist. Our results show that ClusterLink performs
best out of the three tools in scenarios with increased payloads, regardless of the underlying
networking conditions or transmission direction between clusters.  It is closely followed by
Skupper, unless request and reply both transport significant amounts of data.  Finally, when
requesting smaller amounts of data from a service, Submariner slightly outperforms Skupper and
ClusterLink regardless of the inter-node networking conditions.
    
\end{abstract}


\section{Introduction}
With the rapid evolution of mobile networks towards 5G/6G, a myriad of new mobile applications are emerging as ever more distributed in nature. Together with cloud computing in the so-called compute continuum, edge computing is becoming the key solution in mobile networks today to providing flexible, low-latency and distributed computation. To deploy highly distributed applications, container orchestration platforms play a critical role, with Kubernetes not only being the most commonly used solution in the cloud, but also increasingly utilized at the edge. Kubernetes organizes multiple nodes into logical clusters, sharing a common control plane and working in union to deploy and run applications. On the other hand, some specialized environments like mobile or Multi-Access Edge Computing (MEC) might introduce volatile nodes that can leave and join or rejoin clusters, creating coordination and service migration overhead. In edge environments, the restriction of cluster size is often necessary to this end, which in turn limits the application size and complexity that can be deployed per cluster. To overcome this issue, service components have to be spread across multiple clusters while still allowing for communication between them. 

\par The inter-cluster communication in the edge context is in fact in its infancy. In this area, research to date majorly focuses on either the placement of workloads across the infrastructure\cite{bany_taha_efficient_2023}, orchestration of multiple clusters\cite{santos_efficient_2024} or the methods for cluster creation based on various properties, such as node proximity\cite{shahraki_survey_2021}. However, research on how these components can horizontally scale by actually establishing connectivity between them and how this impacts application performance is lacking. Static IP connectivity, while feasible for cloud environments, would require a significant maintenance overhead in dynamic environments like MEC. In fact, such an approach is not suited for large-scale and evolving deployments, therefore necessitating flexible connectivity creation based on service name resolution across cluster boundaries. As we face a rapidly emerging space of open-source projects that address the area of inter-cluster communication, the question arises as to how to assess and evaluate their impact on the application performance in edge environments.


In this work, we experimentally study and comparatively analyze for the first time three notable
open source projects and tools for inter-cluster communication in edge networking and computing:
Submariner, ClusterLink and Skupper. While there are several other tools for general inter-cluster
communication, this study focuses on open-source solutions that do not necessitate sidecar containers
or a specific container network interface in the connected clusters, and as such allow for the
highest degree of compatibility with existing infrastructure. To this end, we engineer an
open-source testbed for inter-cluster communication tools. Its realization uses only
production-grade open-source tools, while it provides the ability to vary the underlying network
conditions to mimic an edge network infrastructure setting with various link qualities. The
inter-cluster communication tools are implemented in a pluggable fashion, thus allowing for a
comparative analysis. We also design a set of innovative case studies for novel distributed mobile
applications, including use cases based on machine learning (ML), which necessitate inter-cluster communication. Finally, we provide measurements and evaluate the impact on application performance. In the measurements, we attempt to answer the following three research questions, which remain unaddressed by existing work for the scope of inter-cluster communication:
\begin{inparaenum}
    \item How does the distance between clusters impact the service response time of latency sensitive applications?
    \item What is the combined effect of cluster distance and payload size on the application's performance?
    \item How do various inter-cluster communication tools comparatively perform for latency sensitive applications?
\end{inparaenum}

The remainder of this article is structured as follows: Section~\ref{sec:related-work} provides the background on the inter-cluster communication tools and the analyzed use cases as well as the related work as state of the art. Section~\ref{sec:system-architecture} presents the system architecture of the testbed used for performance benchmarking. Section \ref{sec:evaluation-setup} describes the design and parameters of the tests executions. Section \ref{sec:results} discusses the results and Section \ref{sec:conclusion} concludes the paper.


\section{Background and Related Work}
\label{sec:related-work}
In this section, we first describe the related work on inter-cluster communication. We then briefly provide the background on Submariner, ClusterLink and Skupper, and finally provide assumptions on application case studies applied in the context of emerging mobile network distributed applications.

\subsection{Related work}

Kubernetes is the de facto default tool for managing workloads in the cloud. It organizes computing
resources in so-called clusters that can be formed based on different characteristics of nodes,
administrative or management goals, or yet a combination of them. Dynamic clusters are formed based
on aspects subject to continuous changes, like node proximity in scenarios with moving compute
nodes~\cite{bany_taha_efficient_2023}. Static clusters are usually based on hardware properties to
form specialized groups, for example, to process tasks that demand hardware accelerators such as
GPUs~\cite{tariq_accelerating_2024}. Such specialized clusters are employed to guarantee the
compatibility of applications that demand specific resources since those devices might leverage
enhancements to utilize specific resources like GPUs\cite{haavisto_unleashing_2022}. Additionally,
scheduling and workload sharing can differ in those particular installations, which could lead to
incompatibility with other Kubernetes clusters~\cite{adufu_application-aware_2024}. General
clustering approaches are analyzed in \cite{shahraki_survey_2021}, showing how dependent the
selection objectives can be on the specific use case.

Distributed application and service function chains can be formed across multiple clusters. The
applications are scheduled and scaled based on node mobility or network properties to minimize
service latency\cite{santos_efficient_2024,ma_mobility-aware_2022}, increase cost-efficiency, reduce
power consumption or boost robustness~\cite{shahraki_survey_2021}. Multi-cluster deployments of
applications are gaining attention because they address a variety of issues faced in distributed
single-cluster environments. On the one hand, there are business reasons to rely on multiple cloud
providers (\ie, multi-cluster environments), such as preventing vendor lock-in and ensuring a global
presence. On the other hand, there are technical reasons, such as the need for edge computing to
fulfill stringent latency requirements of novel applications (\eg extended
reality)~\cite{Charyyev2020}, creating scalability issues. In ~\cite{jeffery_rearchitecting_2021},
the authors show that the Kubernetes control plane can become a bottleneck when a certain cluster
size is exceeded. This restriction limits the size of clusters since consensus has to be maintained
across all member nodes. While the issue can be reduced by changing or improving the used key-value
storage system, this does not provide a complete solution, and the problem can be expected to increase
relative to the geographical spread of the nodes. Additionally, due to the inherent characteristics of edge nodes (\ie constrained computing and storage capacity), some applications can only partially be executed in an edge cluster, which in turn requires communication with other clusters.

In general, research focusing on the inter-cluster communication problem to allow multi-cluster
collaboration is work in progress. In~\cite{Ejaz2024}, the authors perform an assessment of
Submariner's performance as well as the distributed application deployed on top of it with a focus
on CPU utilization. However, this study lacks consideration of two critical dimensions: varying
inter-node/inter-cluster latency and the amount of data being exchanged between the clusters.
In~\cite{Syrigos2023} the authors explore inter-cluster communication by proposing a framework to
deploy a 5G experimental network across distributed infrastructure formed by multiple geographically
spread clusters. The authors evaluated the maximum achievable throughput in inter-cluster
communications provided by Submariner using three different tunneling options (\ie Wireguard, Libreswan, and VXLAN) and three different transport protocols (TCP, UDP and SCTP). While the cited works focus solely on Submariner, we compare the three main tools used for inter-cluster communication in multiple scenarios.

Focusing on cloud-driven scenarios for mobile networks furthermore, paper \cite{Osmani2021} proposed
a solution for multi-cluster management and connectivity employing two open-source tools: KubeFed
and NSM. The Libreswan technology was used to tunnel traffic between clusters. The proposed approach employs TCP, UDP, and SCTP as transport protocols, and the authors benchmark basic data plane performance, collecting maximum throughput and latency metrics between containers. Since latency between nodes
was static, real-world performance in the network edge would diverge significantly from these
findings. A valuable extension of paper \cite{Osmani2021} would in fact be to compare the tool
proposed with some of the production-grade tools that we use in our paper. 

Related work in benchmarking of intra-cluster networking solutions, such as in~\cite{suo_analysis_2018}, focuses on total performance within a single cloud cluster, which is hardware-dependent. 
Paper~\cite{Varghese2021} conducted an extensive survey of benchmarking
performance at the edge and observed that further research is needed to measure the performance of
edge networks efficiently. This survey reveals a gap in explicitly benchmarking orchestrators and
other software that connects the infrastructure into an edge continuum. Indeed, many benchmarks are
limited by physical aspects of edge devices (\eg processing power, chipset throughput, storage, and
memory). Our evaluation differs from previous works as it focuses on the relative response time differences encountered through varying aspects of 1) underlying network, 2)  payload size, and 3)  direction of payload exchange between clusters, instead of absolute metrics like resource utilization.





\subsection{Inter-cluster Communication Tools}
Table~\ref{tab:comparison-of-tools} lists the inter-cluster communication tools that we implement in this
paper: Submariner\footnote{https://submariner.io/} (CNCF, RedHat) and
Skupper\footnote{https://skupper.io/}, along with the emerging
ClusterLink\footnote{https://clusterlink.net/} (IBM)~\cite{Toledo2023}. 
While all three tools tackle the same challenge of inter-cluster service communication, each has unique features that set them apart. 
Skupper, being a layer 7 interconnect, does not need any further
configuration of the CIDR ranges that are used to assign IP addresses to pods or services, making it
scale very well with comparably low configuration effort. Submariner on the other hand works on
layer 3 and employs a broker for connection establishment, leading to only this one component having
to be publicly accessible; it can potentially even be hosted on hardware that is not part of the
communicating clusters. This allows for cluster-to-cluster communication between dynamic endpoints through
NAT as well as firewalls. Using HTTP, ClusterLink shares the benefits of layer 7 connections but
also provides authentication as well as granular permission and access management on workload,
service and cluster level, allowing for a high degree of security and low levels of implicit trust
between endpoints. All three solutions are written in Go, provide state-of-the-art encryption out of
the box and are deployed on top of the respective Kubernetes clusters without the need for
modifications of the individual node's operating systems.

\begin{table}
    \centering
    \caption{Comparison of cluster connectivity tools}
    \label{tab:comparison-of-tools}
    \begin{tabular}{@{}lllll@{}}
        \toprule
        Tool name    & Deployed  & Network    & Protocol  &  Features      \\
                     & version   & driver     &           &                \\
        \midrule
        Skupper.io   & 1.7.3     & mTLS       &   AMQP    & simple configuration \\
        \addlinespace
        Submariner   & 0.18.0    & Libreswan  & IPsec     & NAT compatible \\
                     &           & VXLAN      & VXLAN     &                \\
                     &           & Wireguard  & Wireguard &                \\
        \addlinespace
        ClusterLink  & v0.4.0    & mTLS       & HTTP      & ACLs           \\
        \bottomrule
    \end{tabular}
\end{table}

\subsection{Applications}
The performance evaluation of the inter-cluster communication tools implemented uses key performance indicators (KPIs) and characteristics of emerging mobile computing applications. We choose case studies of two classes of applications: industrial automation and cellular Vehicle-to-Everything (C-V2X). These are illustrative of the real-world verticals that require robust inter-cluster communication and are described as follows.

\subsubsection{Remote Access, Monitoring, and Maintenance}
In an industrial automation setting, numerous sensors installed on production machines distributed
across a factory continuously collect data about the status of machines in the production line. A
massive amount of raw data is cached in a cluster of edge nodes for preprocessing that aggregate the
data where possible. The result of the local processing is sent to another cluster for evaluation
(\eg using machine learning), with the result potentially being used to instruct a control center to
perform the required maintenance~\cite{5g-acia-traffic-model,5g-acia-edge,5g-acia-requirements,3gpp-22.104}.


\subsubsection{Vehicle Decision Assist}
A specific vehicle that is part of a fleet of vehicles platooning on the road decides to overtake
the vehicle in front, for example to optimize fuel consumption, avoid breaking when a preceding
vehicle leaves the platoon at an upcoming exit, or because the vehicle in front is facing a
mechanical problem. Due to road and traffic conditions, the vehicle cannot overtake easily, so the
platoon asks for external information to assist with the maneuver. The exchange of information
happens between the cluster formed by vehicles in the platoon and a cluster of Road Side Units or
oncoming vehicles of the opposite direction (\eg another platoon). The vehicles share speed, length,
and position information to guarantee a safe overtaking. Since reliability and safety is of utmost
importance, the overall system consumes information from multiple systems. In this use case,
multi-cluster communication is required since the platooning vehicles per direction and the roadside
units supporting the vehicles would all be grouped in different clusters~\cite{5gaa-use-cases,5gaa-edge}.




Table~\ref{tab:payload-size} illustrates the extracted measurable KPIs for the applications chosen, including references to the literature that we base these KPIs on. 

\begin{table}
    \centering
    \caption{Average payload size and max. response time per use case}
    \label{tab:payload-size}
    \begin{tabular}{@{}lrr@{}}
        \toprule
        Use case          & Message payload & Max. response   \\
                          & size (byte)     & time (ms) \\
        \midrule
        Remote Access, Monitoring, & 60 - 1000000  & 1000\\
        and Maintenance\cite{5g-acia-traffic-model,5g-acia-edge,5g-acia-requirements,3gpp-22.104}            &                   &         \\
        \addlinespace
        Vehicle Decision Assist\cite{5gaa-use-cases,5gaa-edge}    & 1000              & 100       \\
        \bottomrule
    \end{tabular}
\end{table}

\section{System Architecture and Testbed Design}
\label{sec:system-architecture}

\begin{figure}[htbp]
    \centering
    \includegraphics[width=1.0\linewidth]{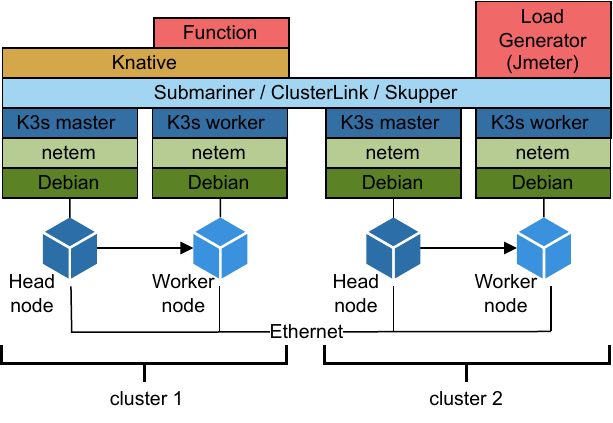}
    \caption{Testbed design}
    \label{fig:testbed-design}
\end{figure}

Fig.~\ref{fig:testbed-design} illustrates our testbed. It includes four physical machines with the specifications listed in Table~\ref{tab:hardware}, all of them are equipped with Gigabit interfaces and connected through a single Gigabit Ethernet switch. They represent off-the-shelf, but well-equipped consumer hardware which we assume to roughly represent the computational capacity of power-restricted ($<=$ 100 Watts) edge nodes with small form factors that could be deployed in constrained spaces, e.g., in curb cabinets, cars or industrial machines. 

\begin{table}[htbp]
    \centering
    \caption{Testbed Hardware}
    \label{tab:hardware}
    \begin{tabular}{@{}lllc@{}}
        \toprule
        Cluster & Role & CPU & Memory \\
        \midrule
        cluster 1 & headnode & Intel Core i5-7600 @ 3.5 GHz & 16 GB  \\
                  & worker   & Intel Core i5-10600 @ 3.3 GHz & 16 GB \\
        \addlinespace
        cluster 2 & headnode & Intel Core i5-7600 @ 3.5 GHz & 32 GB  \\
                  & worker   & Intel Core i7-6700 @ 3.4 GHz & 32 GB  \\
        \bottomrule
    \end{tabular}
\end{table}

The software stack used consists of hierarchical layers, as seen in Fig.~\ref{fig:testbed-design}.
Since we publish our code base on Github\cite{code-base}, we design the testbed exclusively around open-source tools, akin to \cite{Carpio2022}. 
The functionalities provided by each layer are described below. This environment guarantees a common baseline for all three tools. Most notably, our testbed allows for the manipulation of network properties like delay, jitter, and packet loss through software, while keeping all other factors static, thus enabling comparative performance measurements. 

\subsection{Operating System and Kubernetes}
We deploy a linux-based operating system to provide basic functionality as well as all prerequisites needed by the container orchestrator. Here, we chose Debian bookworm due to its server-grade stability and wide acceptance, which makes it one of the most tested platforms for edge and cloud tools. We also employ the tool \emph{netem} for manipulation of the Ethernet interface by artificially delaying or dropping frames, therefore emulating delay, jitter and packet loss as experienced relative to geographical distance or connection quality.

\par To provide a container orchestration layer, we choose the k3s Kubernetes distribution meant to
be used for lightweight and edge devices, v1.30.3+k3s1. It provides the majority of features
available in the regular Kubernetes version while having a much smaller footprint. Two clusters are
formed with one worker and one headnode each and a total of four different IP ranges between the two
clusters for their respective services and pods. As can be seen in Fig.~\ref{fig:testbed-design},
the worker nodes run the application and load generator respectively, while the headnodes are used
exclusively to run each cluster's control plane. This separation ensures that the impact of control
plane tasks on the workers is minimized, while also replicating real-world use cases where at least
one node per cluster has to be available via a defined address to allow for management and
negotiation of connection endpoints.

\subsection{Inter-Cluster Communication}
Two clusters are connected through the inter-cluster communication tool under test, namely Submariner, ClusterLink, and Skupper, as shown in the light blue box on Fig.~\ref{fig:testbed-design}. This setup allows services in cluster 2 to access exposed services in cluster 1. The inter-cluster communication tools are deployed as Kubernetes \textit{yaml} definitions and are responsible for service discovery, service address resolution, and data transport across the connected clusters.

\subsection{Applications}
The three example applications are functions served on top of the Knative\cite{knative} framework, which handles requests and provides HTTP endpoints. 
Their source code is published on GitHub\cite{functions}. 
While all of those applications are deployed at the same time, we interact with them sequentially. 
Automatic scaling in Kubernetes is deactivated since we do not aim to measure total achievable performance on the given hardware but instead relative response time differences. 
The functionality of each deployed function can be described as follows:

\subsubsection{payload-echo}
This function accepts any payload and echoes the same message back to the sender by setting the
response body to that received payload. It is used in testing our \emph{Remote Maintenance
(ML Inference)} use case, since it emulates the transfer of aggregated data from one cluster to
another with an equally sized response. In the test, the result shall be transmitted back to the
sending cluster, which is the behavior that this function exhibits.

\subsubsection{payload-recv}
This function receives any payload sent as HTTP body and acknowledges it with an HTTP 200
response without body. It is used for the \emph{Remote Maintenance (ML Training)} application, since we can see the same data transfer here as in the previous function with the difference that the data is not transferred back. This behavior would be expected whenever an ML model is trained, since the sending cluster would not have to wait for the training to complete.

\subsubsection{static-reply}
This function simply replies to any requests with a static message, regardless of the request body.
We use it to mimic the \emph{Vehicle Decision Assist} application, where payload data is only received and not sent. Here, one cluster queries the information from another cluster without submitting any payload itself.

\subsection{Load Generator}
To generate requests to the application, we utilize Jmeter in version 5.5, a load testing tool that allows for generating parallel and sequential HTTP requests. We deploy the tool as a container on top of Kubernetes to emulate the communication between two containerized services deployed on different clusters. The tool measures the time it takes between sending the request and receiving the reply and saves it to a csv file.

\section{Evaluation Setup}
\label{sec:evaluation-setup}

The main challenge in the evaluation setup is to create the testing environment and applications as
close as possible to the real world use case such that meaningful results for the application response times can be obtained. To this end, and based on the description provided in Section~\ref{sec:related-work}, we map the analyzed communication patterns to our example applications, and design the matching Jmeter tests to evaluate them. We then gather the response times achieved through the different tools across a set of six different infrastructure scenarios that mimic edge environments with different characteristics.
To create a baseline, we start with a test without any interface manipulation, leaving only the delay, jitter and packet loss experienced physically by the transmission. This establishes a reference value for each tool when running on our provided hardware and software stack. 

We then start introducing delay, jitter and packet loss to the frames transmitted through the Ethernet interface, emulating a transmission over higher distances or lower quality connection media between all nodes as listed in Table~\ref{tab:link-parameters}. The naming of each parameter set is based on the additional delay introduced to an average round trip time between any two nodes; \eg 10ms in edg10. We base our values on the measurements described by the authors in \cite{Charyyev2020} and extrapolate them to gain a set of test values.

Each Jmeter test is configured with base parameters like the number of sequential requests (40), the
number of parallel threads (1) and the reuse of TCP sessions (enabled). Every test is executed
across all six scenarios and repeated ten times on each of them to guarantee reproducible results. Before every run, the whole infrastructure is reverted to a clean, baseline state to prevent cross-contamination of the results.

\begin{table}[htbp]
    \centering
    \caption{Emulated link parameters per scenario}
    \label{tab:link-parameters}
    \begin{tabular}{@{}lrrr@{}}
        \toprule
        Scenario     & Delay & Jitter & Packet Loss   \\
        \midrule
        lcl          & +0 ms    & +0 ms    & +0\%     \\
        edg2.5       & +1.25 ms & +0.25 ms & +0.02\%  \\
        edg5         & +2.5 ms  & +0.5 ms  & +0.04\%  \\
        edg10        & +5 ms    & +1 ms    & +0.08\%  \\
        edg15        & +7.5 ms  & +1.5 ms  & +0.12\%  \\
        edg20        & +10 ms   & +2 ms    & +0.16\%  \\
        \bottomrule
    \end{tabular}
\end{table}

\subsection{Applications}

All tests are conducted based on Fig.~\ref{fig:data-exchange}, including the request and response
direction. Table~\ref{tab:data-exchange} details the payload sizes of the files used to generate the requests/responses.

\subsubsection{Remote maintenance inference}
\begin{figure}[htbp]
    \centerline{\includegraphics[width=0.35\textwidth]{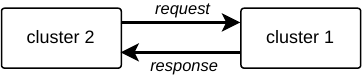}}
    \caption{Data exchange across clusters}
    \label{fig:data-exchange}
    \end{figure}

In this class of applications, the payload-echo function emulates the transmission of aggregated
data to a specialized cluster for processing. The destination cluster communicates back to the
aggregation cluster. Therefore, data has to be transmitted in both directions, where, depending on
the specific data and actuators, the amount per direction could fluctuate significantly. For
simplification, our destination cluster echoes back the received data, which should approximate the system behavior described in the use case. In Fig.~\ref{fig:data-exchange}, the payload size of the request and the response are therefore identical. During the execution of this test, the KPI for the response time is considered to be 1000 ms. Once this cannot be ensured by the different tools, potentially catastrophic failure in a production system can be assumed.

\begin{table}[h]
    \centering
    \caption{Payload size of data exchange between clusters}
    \label{tab:data-exchange}
    \begin{tabular}{@{}lll@{}}
        \toprule
        Use case     & \multicolumn{2}{c}{HTTP Payload size} \\
                     & \textit{request} & \textit{response} \\
        \midrule
        Remote maintenance & 60B, 1KB, 2KB, 10KB, & 60B, 1KB, 2KB, 10KB, \\
        inference          & 100KB, and 1MB       & 100KB, and 1MB       \\
        \addlinespace
        Remote maintenance & 60B, 1KB, 2KB, 10KB, & 0B                   \\
        training           & 100KB, and 1MB       &                      \\
        \addlinespace
        Vehicle decision   & 0B                   & 1000B                \\
        assist             &                      &                      \\
        \bottomrule
    \end{tabular}
\end{table}

\subsubsection{Remote maintenance (ML training)}
In this test, using the function payload-recv, we emulate the behavior of the system sending data
for training an ML model. The data is assumed to have been aggregated, resulting in variable-sized
data in a cluster that is sent to a second cluster. Since the data is used for ML training, the
destination cluster, \textit{cluster 1} in Fig.~\ref{fig:data-exchange}, only acknowledges
receiving the data without sending any payload. The payload of the request ranges from 60B to 1MB,
while the payload of the response is zero bytes (0B) as detailed in Table~\ref{tab:data-exchange}.
Since, depending on the application, machine learning training can have high tolerance regarding the response time we choose an arbitrary maximum response time of 1000 ms. This allows us to compare the effect of a simple acknowledgement to the extensive application response of the previous test.

\subsubsection{Vehicle decision assist}


Here, we use the static-reply function. A car in a platoon is overtaking and coordinating with
either the roadside units of the section, or alternatively/complementary an oncoming platoon of
vehicles. It has to receive data from the other cluster, \eg platoon length, speed, and position, with low tolerance for delay. Therefore, we assume an empty payload for the request, which aims to trigger a response with a fixed payload size of 1000B, as shown in Table~\ref{tab:data-exchange}.

As in all tests, we evaluate the results to determine with which inter-cluster network parameter set
the response times break the maximum value to ensure save operation. Here, we set this value to 100
ms before the information is deemed so outdated that it loses its meaning and endangers safe
operation of the machines and safe transportation of the passengers.

\section{Measurements and Results}
\label{sec:results}

In this section, we present the measured values and the results achieved through the tests described previously.  Our measured application response times are defined as, \ie 
\begin{equation*}
    T_{total} = 
    T_{reqSend} +
    T_{hostProc} + 
    T_{appProc} + 
    T_{replySend}
\end{equation*}
where $T_{reqSend}$ is the request sending time including possible retransmissions, $T_{hostProc}$
is the host processing time at the destination cluster which includes the time taken by
(de-) fragmentation and reassembly of the request. $T_{appProc}$ is the delay introduced by the
processing of the application including potential processing of the payload, which depends on the
parameter size in bytes. In practice, the differences introduced by the processing of different
parameter sizes in our experiments are compensated for by the baseline measurements (loc0 scenario) in our results. Finally, $T_{replySend}$ is the reply sending time including possible delays introduced by retransmission mechanisms in the network layer.


During the evaluation, we focus especially on the two sending times since they are specific to the inter-cluster communication tools and point out differences between them per tool.

\subsection{Remote Maintenance (ML Inference)}
\begin{figure}[htbp]
\centerline{\includegraphics[width=0.5\textwidth]{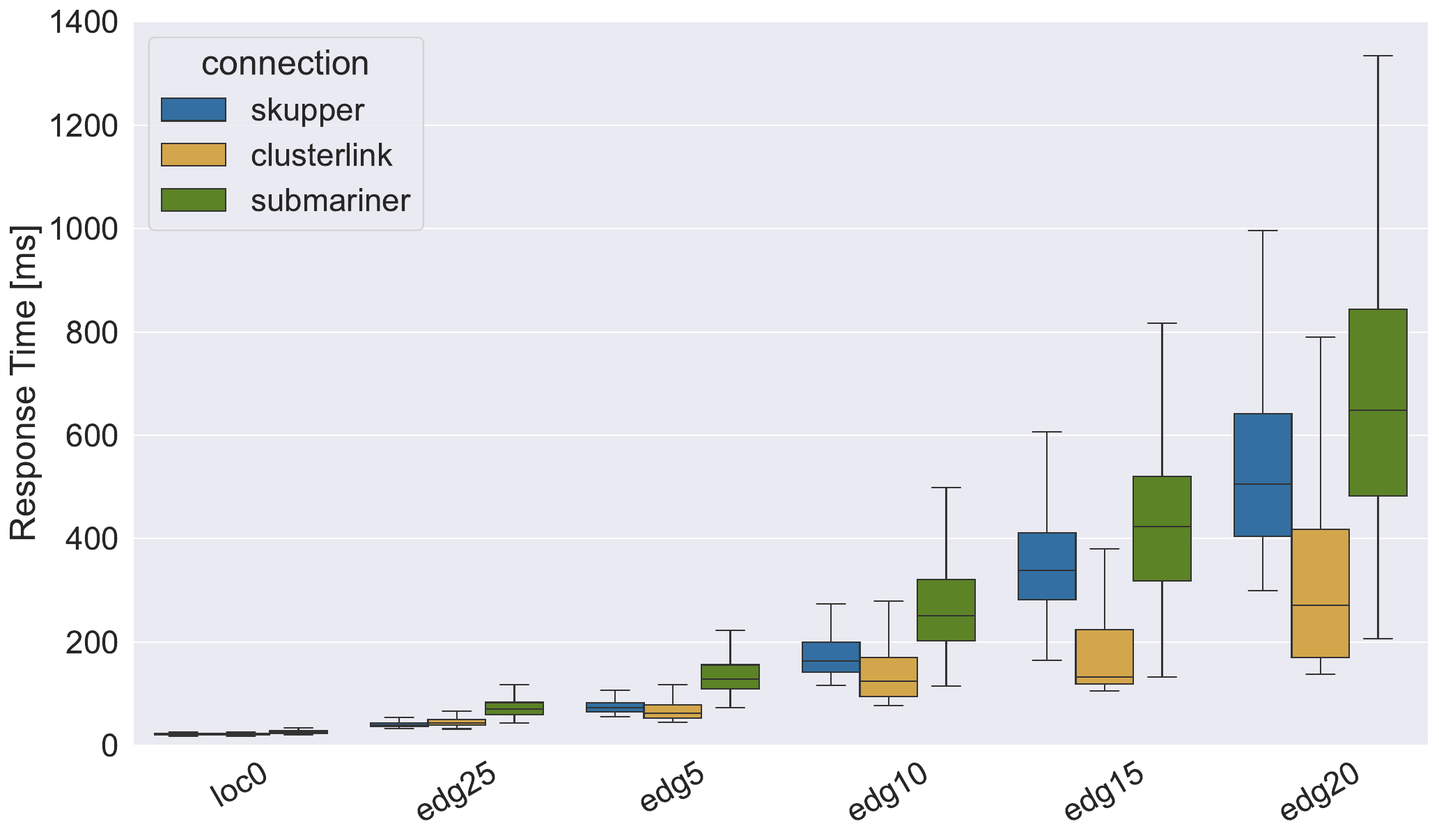}}
\caption{Results for Remote Maintenance (Inference) 1MB}
\label{fig:res-echo-1m}
\end{figure}

Fig. \ref{fig:res-echo-1m} shows the response times achieved via the respective inter-cluster
communication tools for the ML inference in the Remote Maintenance use case with a payload of 1MB, the
largest one in our value set. 
In this case study, the data sent is echoed back to the sender, generating a response of the same size as the request.
We can observe a continuous, exponential increase of response times relative to the deteriorating
networking conditions when using Submariner and Skupper, with Skupper achieving roughly two thirds
of the response time of Submariner in our worst case. 
While Clusterlink is subject to the same trend, it seems to better deal with the additional delay
imposed on the network connection between clusters, showing response times almost 50\% lower than
Skupper when using our lowest-quality links. 
In general, all tools were able to handle the request and response without their median response
times exceeding the 1000 ms limit for the use case as defined in Table~\ref{tab:payload-size}. 
Submariner however occasionally exceeds the tolerable upper bound slightly when the parameter set with 20 ms of
delay is introduced to the connection, which indicates that the tool is more sensible to link delay,
delay variance and packet loss when compared to other tools. 


\begin{figure}[htbp]
\centerline{\includegraphics[width=0.5\textwidth]{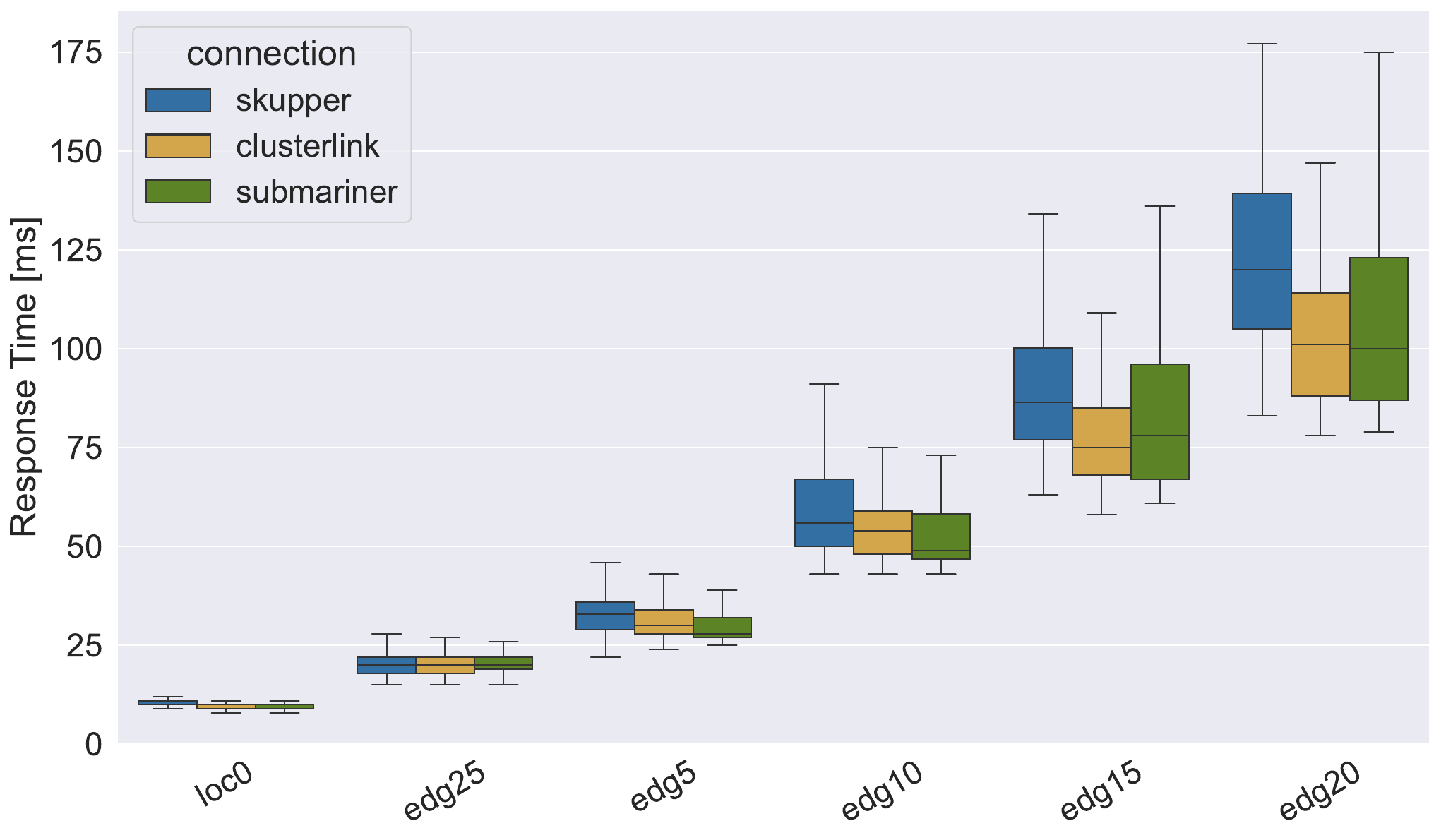}}
\caption{Results for Remote Maintenance (Inference) 100KB}
\label{fig:res-echo-100k}
\end{figure}

For smaller payloads, ClusterLink and Submariner perform similarly until the eg10 value set, with
Submariner being slightly outperformed by ClusterLink in most cases as depicted in
Fig.~\ref{fig:res-echo-100k}.  However, the difference is subtle and in consideration of run to run
variance, their performance can be considered equal here.
Skupper, while initially delivering comparable performance to the other tools breaks away in the
edg15 scenario by delivering even higher response times, by almost 20\% in the edg20 scenario,
indicating a downside of using AMQP in this environment. From a response time perspective, this makes
Skupper the last choice out of the three tools for inter-cluster communication when dealing with
applications that need to send medium-sized payloads to a service that in turn replies with small payloads.


\begin{figure}[htbp]
\centerline{\includegraphics[width=0.47\textwidth]{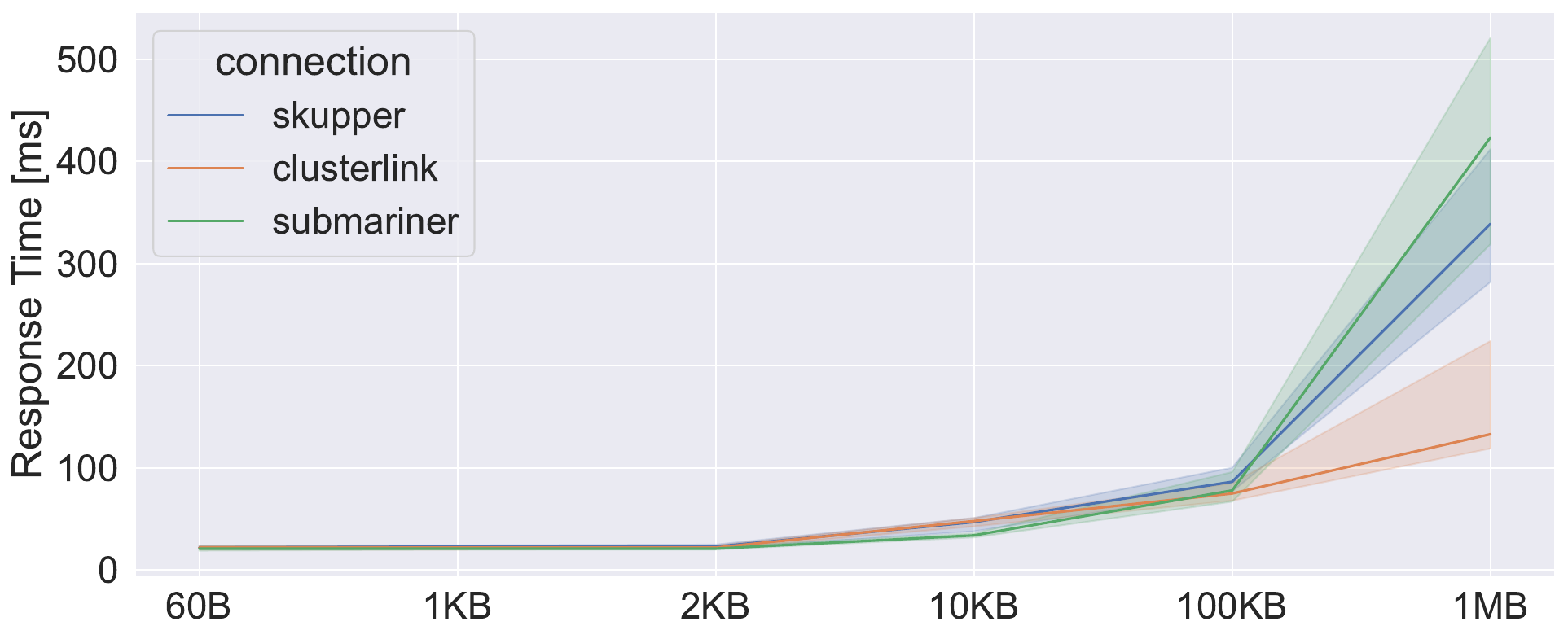}}
\caption{Results for Remote Maintenance (inference) edg15}
\label{fig:res-echo-edg15}
\end{figure}

In Fig.~\ref{fig:res-echo-edg15}, we analyze the payload size's impact on the tools' performance for
the remote maintenance inference use case in the scenario with the worst conditions
that still allow all tools to perform below the boundary; edg15. We can observe similar performance
for the three tools, with mean values below 100 ms for payloads up to 100 KB.
With the payload size increased to 1MB, we see how the performance of the tools
diverges, with drastic increases of the response times of Submariner, Skupper
scaling slightly better, while ClusterLink's jump in response times is significantly less pronounced.

We can therefore conclude that for workloads where small amounts of data are sent to remote
services that reply with similar amounts of data, the three tools perform similarly to each other.
For medium and larger data, i.e. 100 KB and more, the tool selection becomes a decision of significant impact. 

\subsection{Remote Maintenance (ML Training)}
\begin{figure}[tbp]
\centerline{\includegraphics[width=0.5\textwidth]{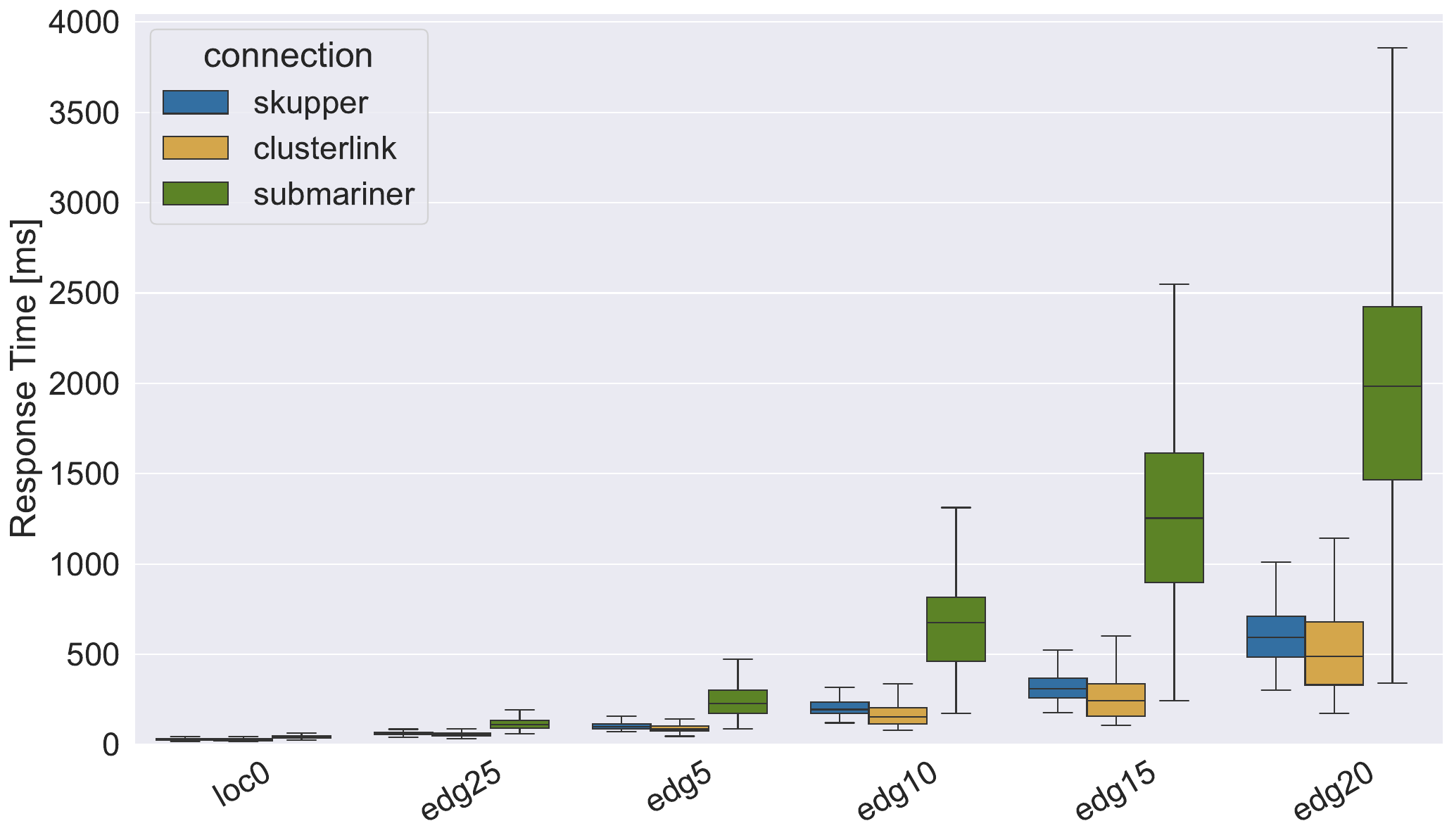}}
\caption{Results for Remote Maintenance (Training) 1MB}
\label{fig:res-recv-1m}
\end{figure}

Fig.~\ref{fig:res-recv-1m} shows the results of the ML training case studies for 1MB payload size,
which, differently from the inference use case, does not consider communication back to the request
source besides acknowledgments. We can observe ClusterLink and Skupper behaving similarly to
each other with slightly lower medians for ClusterLink, while it now also exhibits a slightly higher
spread. Similarly to the ML inference for the remote maintenance use case with payloads of 1 MB,
Submariner performs worst among the tested tools. Here, it presents about four times the median
values of ClusterLink and Skupper in the edg20 scenario along with a high standard deviation in the
observed data. In contrast to the previous use case, Submariner exceeds the boundary of 1000 ms 
already in the edg10 scenario with some measurements and in the edg15 scenario with its median
value. ClusterLink and Skupper however remain below the threshold with their median values but
slightly exceed the boundary in the edg20 scenario with a small portion of the observed samples.

\begin{figure}[tbp]
\centerline{\includegraphics[width=0.5\textwidth]{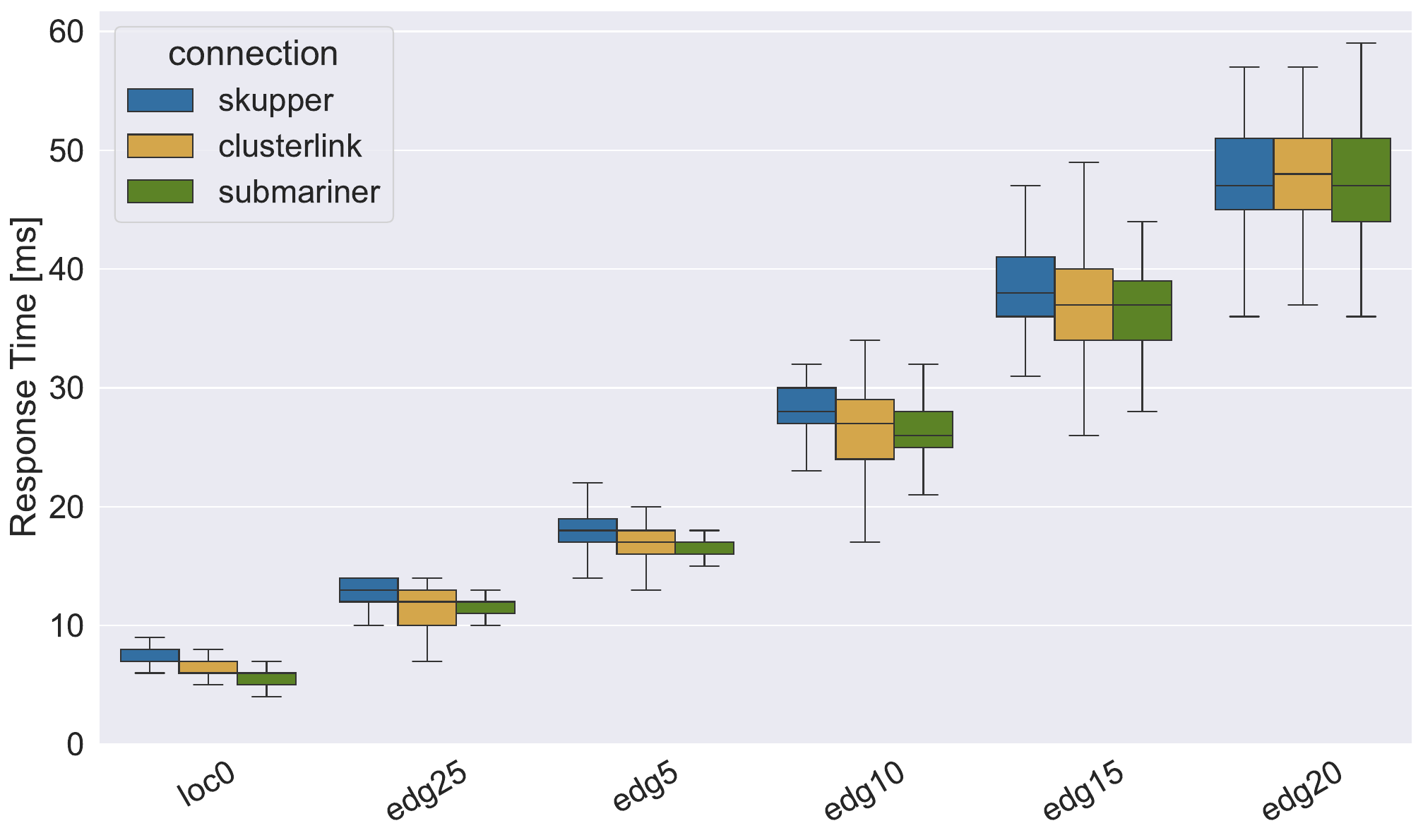}}
\caption{Results for Remote Maintenance (Training) 60B}
\label{fig:res-recv-60b}
\end{figure}

The results for experiments with smaller payloads of 60 Bytes are presented in
Fig.~\ref{fig:res-recv-60b}. The plots show almost identical performance for the three inter-cluster
communication tools, to the level of some even reaching identical median values in the edg15 and edg20
scenario.

\begin{figure}[tp]
\centerline{\includegraphics[width=0.47\textwidth]{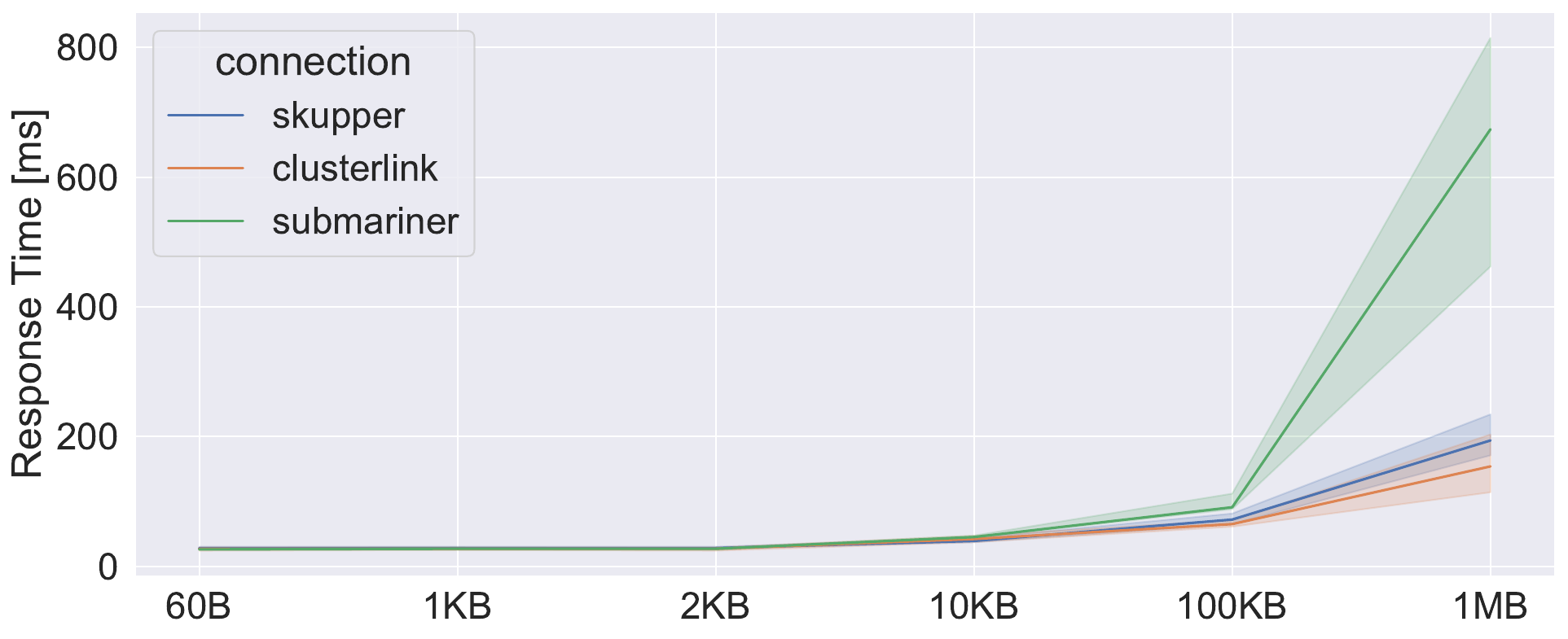}}
\caption{Results for Remote Maintenance (Training) edg15}
\label{fig:res-recv-edg10}
\end{figure}

In Fig.~\ref{fig:res-recv-edg10}, we show the impact of varying payload sizes on the tools'
performance for the remote maintenance training use case under the worst networking conditions that still
allow all tools to achieve median response times below the boundaries. We can observe similar
performance for all three tools until the payload size exceeds 2 KB. Higher payloads lead to
Submariner's performance degrading significantly quicker than for ClusterLink and Skupper in
relation to the payload size, indicating a clear advantage of the mTLS-based implementations over
the IPsec-based Submariner due to less overhead of this approach. ClusterLink then brings this
overhead down further by utilizing HTTP instead of AMQP for transport, resulting in even better
performance than Skupper.

Curiously, we observe the worst performance of Submariner when data is not sent back to
the tester container (or client), apart from acknowledgments. We can therefore conclude, that
Submariner's implementation seems to have suboptimal handling of HTTP POST responses when it comes
to larger payloads, which makes it unsuitable for this application when network latency is above
7.5 ms in each direction (assuming a symmetric link, therefore 15 ms for round trip). 

\subsection{Vehicle Decision Assist}
\begin{figure}[htbp]
\centerline{\includegraphics[width=0.5\textwidth]{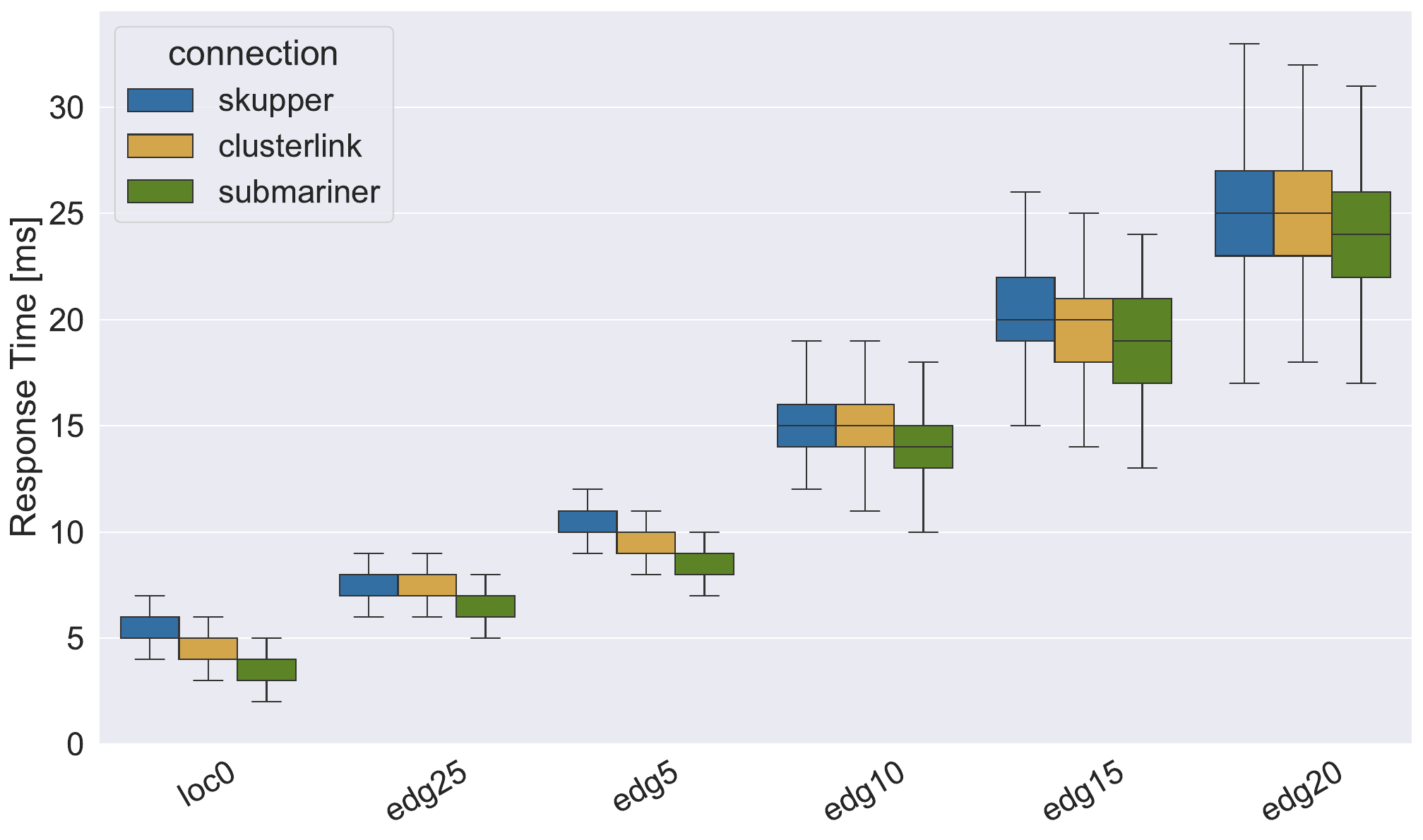}}
\caption{Results for Vehicle Decision Assist}
\label{fig:res-hw}
\end{figure}

Fig.~\ref{fig:res-hw} shows the results for the vehicle decision assist class of applications.
ClusterLink and Skupper again present very similar performance with their median values being
identical across almost all scenarios. Submariner however seems to consistently perform slightly
better (1 ms lower median response times) than the other tools. 
When handling these small payloads, according to our results, the inter-node connection quality only
has a linear influence on the total response time. The response times of all three tools remain
inside the set boundary of 100 ms, making each of them a suitable choice for this class of
application.
A further increase of the distance between the nodes beyond a round trip time of 20 ms would be
unrealistic, since such a high latency for devices that are in danger of physically touching each
other is highly unlikely.


\vspace{-0.2em}


\subsection{Discussion}
Since this is an experimental research work, the applicability of our results is limited in the following ways:

\subsubsection{Hardware}
Our measurements strongly depend on the individual hardware, and we cannot completely exclude the possibility that the three tools would score in a different order on different hardware due to better encoding or encryption performance. We however believe, that the tested hardware represents a reasonably broad group of edge devices and therefore should be able to provide a general idea of which tools suit which circumstances.

\subsubsection{Protocols}
The results might not apply to streaming protocols used by real-time applications. With their way of data partitioning and transmission, they should be evaluated separately.

\subsubsection{TCP Connection Reuse}
Our tests have had the ``reuse TCP connections'' activated in JMeter which we believe is a reasonable approach for traffic between two clusters. This however does not represent classes of applications where multiple clusters or services communicate with a common destination cluster. Individual results for such circumstances can therefore differ from what we have presented here.

\subsubsection{Payload Size}
The step increase of our payloads is quite large to allow for displaying a wide variety of different values. This might or might not reflect the performance to be expected with other applications and therefore should be repeated with values closer to the intended data sizes.

\subsubsection{Tool Maturity}
At the time of testing, Clusterlink is available in v0.4.0, meaning it has not yet hit a stable release target. Through code improvements, its performance might be subject to change, while Submariner and Skupper are established and production-ready tools.

\subsubsection{Inter-node and Inter-cluster Latency}
In this work, we assume equal distance between two nodes of the same cluster and between different clusters. In reality, the inter-cluster distance would often be higher than the distance within clusters, depending on the metric that the cluster is formed on.

\subsubsection{Submariner Network Driver}
We only explored Submariner’s libreswan (IPsec) driver, while other research \cite{Syrigos2023} has
shown that its performance could vary when different drivers are used.

\section{Conclusion}
\label{sec:conclusion}
   In this work, we evaluated for the first time the impact of inter-cluster communication on edge computing performance by using three prominent, open source inter-cluster communication projects and tools, i.e., Submariner, ClusterLink and Skupper. We developed an open source testbed that integrates these tools in a modular fashion, and experimentally benchmarked the sample applications, including the ML class of applications, on their performance under varying networking conditions.
   
   \par  In answering the research questions stated upfront, we proved that the response times have a linear relation to the geographical distance between all nodes when the transmitted data is low in size, while it can approach exponential growth with large payload sizes.
   We also observed that the combined impact of geographical spreading of nodes and increasing payload size on
   the considered applications leads to exponential growth of application response times, at least
   for data being transferred into both directions and for data being sent. For the three tools
       analyzed, we found that ClusterLink outperformed Submariner and Skupper whenever larger data was
       exchanged with a service, Submariner performed slightly better for small payloads that are
       received via HTTP GET requests and Skupper performed decently unless medium-sized payloads
       had to be pushed via HTTP POST requests.

   \par In more general terms, we can say that for inter-cluster communication, creating network
   layer tunnels like Submariner seems to provide higher efficiency for small payloads but
   suffers when dealing with large amounts of data as soon as significant delay, variance and packet
   loss is experienced on the connecting link. Solutions with the ability of proxying TCP
   connections are at advantage here, delivering lower overall response times. Furthermore, use
   cases where data has to be sent to another service with significantly sized replies seem to
   suffer when using Skupper or Submariner in edge networks with suboptimal inter-node connections,
   showing that using HTTP for traffic transport leads to better performance.

   \par It should be noted that the individual choice of software for a real-world setting should
   however always evaluate not just the performance, but also the feature set of the tools. For
   instance, if strong security is needed, ClusterLink's authentication and access control might
   prevail, just as Submariner's flexibility for connection establishment might make it more a more
   attractive choice. Also, when simplicity of the configuration setup is the main concern, Skupper
   might be chosen as the simplest solution.

\section*{Acknowledgment}
This work was partially supported by the German Federal Ministry of Education and Research (BMBF) project 6G-ANNA, grant agreement number 16KISK100, and the project EU HORIZON ICOS under GA no. 101070177.

\bibliographystyle{IEEEtran}
\bibliography{refs}

\end{document}